\documentclass[prl,twocolumn,showpacs,preprintnumbers,amsmath,amssymb,superscriptaddress]{revtex4-1}

\usepackage{graphicx}
\usepackage{dcolumn}
\usepackage{bm}
\usepackage{color}

\begin{document}

\title{Enhanced thermoelectric coupling near electronic phase transition: the r\^ole of fluctuation Cooper pairs}

\author{Henni Ouerdane}\email{henni\_ouerdane@hotmail.com}
\affiliation{Russian Quantum Center, 100 Novaya Street, Skolkovo, Moscow Region 143025, Russia}
\affiliation{Laboratoire Interdisciplinaire des Energies de Demain (LIED) UMR 8236 Universit\'e Paris Diderot CNRS, 4 Rue Elsa Morante, 75013 Paris France}
\author{Andrey A. Varlamov}
\affiliation{Russian Quantum Center, 100 Novaya Street, Skolkovo, Moscow Region 143025, Russia}
\affiliation{CNR-SPIN, Viale del Politecnico 1, I-00133, Rome, Italy}
\author{Alexey V. Kavokin}
\affiliation{Russian Quantum Center, 100 Novaya Street, Skolkovo, Moscow Region 143025, Russia}
\affiliation{CNR-SPIN, Viale del Politecnico 1, I-00133, Rome, Italy}
\author{Christophe Goupil}
\affiliation{Laboratoire Interdisciplinaire des Energies de Demain (LIED) UMR 8236 Universit\'e Paris Diderot CNRS, 4 Rue Elsa Morante, 75013 Paris France}
\author{Cronin B. Vining}
\affiliation{ZT Services 1685 H Street, \#872, Blaine, WA USA 98230}

\date{\today}

\begin{abstract}
Thermoelectric energy conversion is a direct but low-efficiency process, which precludes the development of long-awaited wide-scale applications. As a breakthrough permitting a drastic performance increase is seemingly out of reach, we fully reconsider the problem of thermoelectric coupling enhancement. The corner stone of our approach is the observation that heat engines are particularly efficient when their operation involves a phase transition of their working fluid. We derive and compute the thermoelastic coefficients of various systems, including Bose and Fermi gases, and fluctuation Cooper pairs. Combination of these coefficients yields the definition of the thermodynamic figure of merit, the divergence of which at \emph{finite} temperature indicates that conditions are fulfilled for the best possible use of the thermoelectric working fluid. Here, this situation occurs in the fluctuation regime only, as a consequence of the increased compressibility of the working fluid near its phase transition. Our results and analysis clearly show that efforts in the field of thermoelectricity can now be productively directed towards systems where electronic phase transitions are possible. 
\end{abstract}


\maketitle

\paragraph*{Introduction and rationale}Thermoelectric phenomena in conductors emerge from the fundamental coupling between the energy and the electric charge that each mobile electron carries, and they manifest themselves as coupled transport of heat and electricity. In a thermodynamic picture, thermoelectric devices operating as generators or refrigerators, are heat engines where the conduction electrons act as the working fluid. Hence, as it is purely electronic in nature, the thermoelectric conversion is a direct process, which is not system size-dependent and does not entail complex dissipative mechanisms owing to the absence moving parts. Thermoelectric systems may thus present strong advantages over traditional heat engines but the energy conversion efficiency still is far too low to envisage wide-scale applications in the near future \cite{Vining2009}. 

Interest in thermoelectricity has much varied since its early days in the 19th century \cite{pioneer1,pioneer2,pioneer3}, showing a particular surge concomitant of the fast progress in semiconductor physics in the 1950's and 1960's and the ensuing improvement of thermoelectric device performance, which is assessed against efficiency or coefficient of performance depending on the operating mode. It has become customary to relate either of these latter to the dimensionless figure of merit $ZT$, which combines the materials transport coefficients, namely the Seebeck coefficient $s$, the electrical conductivity $\sigma$, the thermal conductivity $\kappa$, and the average temperature $T$ across the system \cite{Ioffe}:
\begin{equation}\label{ZTe}
ZT = \frac{\sigma s^2}{\kappa}~T = \frac{s^2}{L\left(1+\kappa_{\rm lat}/\kappa_{\rm e}\right)}
\end{equation}
\noindent where $\kappa$ entails both electron and lattice thermal conductivities, $\kappa_{\rm e}$ and $\kappa_{\rm lat}$, with $\kappa = \kappa_{\rm e}+\kappa_{\rm lat}$, and  $L=\kappa_{\rm e}/\sigma T$ is the Lorenz number. The ratio $L$ may be viewed as a quantitative measure of a system's relative ability to conduct heat with respect to its ability to conduct electrical charges. To envisage applications for thermoelectric systems other than those for which sustainability and reliability are more important than low-level efficiency and high cost, values of $ZT$ greater than 4 are mandatory \cite{Vining2009}.

Although they were seen as very promising candidates for the development of thermoelectric applications, bulk semiconductors did not prove to be the long-awaited miracle materials: Goldsmid\cite{Goldsmid} anticipated that $ZT$ would not easily reach 1, and that one could hardly hope that it will ever go beyond 2. The 1990's works of Dresselhaus and co-workers on transport in low-dimensional thermoelectric systems \cite{Hicks1,Hicks2,Hicks3} inspired bandstructure engineering, particularly focused on the effective mass, as a way to increase $ZT$. So far, these efforts have had limited success so attention turned to lowering $\kappa_{\rm lat}$ since thermal energy transferred by phonons simply represents a useless heat leak. The paradigm in thermoelectric materials then became Slack's so-called electron crystal - phonon glass system \cite{Slack}, which provided a strong impetus for the continued development of the field of thermoelectricity, which to date remains as active as ever \cite{bookTE1,bookTE2}.

Recent progress in materials science and nanostructure engineering \cite{MaterRev1,MaterRev2,MaterRev3,MaterRev4} essentially entails lowering of lattice heat conduction and enhancement of the Seebeck coefficient of thermoelectric materials, but even ``promising'' materials do not, as of yet, boast the much sought-after minimal requirements, and they may also suffer from a variety of practical problems that preclude wide-scale applications \cite{Dennler}. Further, it has become obvious that, despite the tremendous progress in nanostructure engineering over the last twenty years, one cannot lower the lattice thermal conductivity arbitrarily, and not even below a threshold that would permit obtainment of at least $ZT = 3$. Knowing that the Lorenz number is roughly the same across a wide range of metals, and that the transport coefficients may not vary drastically under standard working conditions, explain to some extent the difficulty to obtain $ZT\gg 1$. So we may turn to an aspect of thermoelectricity, which is always neglected: the thermodynamics of the electronic working fluid and its ability to transport entropy; but this is not so simple since $\kappa_{\rm e}$ relates to two phenomena: heat transfer by conduction (described by Fourier's law) and heat transfer by electron convection \cite{ApertetConv}. The former is constrained by the Wiedemann-Franz law, which applies for metals and degenerate semiconductors; the latter represents the actual thermoelectric flux and hence it can only be seen as the ``useful'' contribution to heat transfer across the thermoelectric system submitted to a temperature bias. 

Enhancing the thermoelectric convective flux may be done by increasing the temperature difference across the system, but this is not a satisfactory solution: though loss through electronic thermal conduction may be largely compensated by the competing convective process, heat transfer by lattice conduction and hence inefficiency also increases. In a context very different from thermoelectricity, the analysis of the effects of phase transitions on mantle convection \cite{Christensen}, showed that if the phase transition is exothermic, the release of latent heat enhances convection. In general, regimes where convection dominates the heat transfer process are characterized by high values of the Prandtl number, $\Pi$, which may be simply defined as follows \cite{Blundell}:
\begin{equation}
\Pi = \frac{\nu}{D} = \frac{\eta/\rho}{\kappa/(\rho C_P)}
\end{equation}
\noindent where $\nu$ is the kinematic viscosity, $D$ is the thermal diffusivity, $\eta$ is the viscosity, $\rho$ is the density, and $C_P$ is the heat capacity at constant pressure of the considered fluid. In simple models the thermal conductivity is given by $\kappa =\eta C_V$, so one sees that the Prandtl number is proportional to the heat capacity ratio $C_P/C_V$, also known as the isentropic expansion factor $\gamma$. The Prandtl number thus provides a link between the thermodynamic properties of the fluid and its capacity for convective heat transfer.

In the present work, we are interested in the properties of the thermoelectric working fluid whose specific study is always neglected in favour of the engine itself in a broad sense (i.e., materials, system configuration, structure). We show that the most profitable conditions for the convective thermoelectric transport are those which bring the working fluid near a phase transition. With the relevant thermoelastic coefficients, we define the figure of merit $Z_{\rm th}$, which is the thermodynamic counterpart of $Z$. Note that $Z_{\rm th}$, as a property of the working fluid, does not contain the lattice term $\kappa_{\rm lat}$. Clearly, we do not aim to propose at present an actual system boasting values of $ZT$ much greater than the highest ones achieved so far; we primarily aim to show that efforts must concentrate on electronic systems that may undergo a phase transition for performance enhancement, and to provide also insight into the fundamental difficulty to increase $ZT$. We thus consider the regime where the so-called superconducting fluctuations \cite{Larkin} above the critical temperature $T_{\rm c}$ appear: while still in its normal phase, the electronic system boasts some particular effects, which pertain to the superconduting phase. Indeed, with the presence of fluctuation Cooper pairs, properties such as, e.g., conductivity and heat capacity, increase significantly as the system approaches the critical point. Fluctuation Cooper pairs play a major r\^ole in the thermoelectric properties of high-$T_{\rm c}$ superconductors, in relation to their transverse thermoelectric response to an applied thermal gradient (Nernst signal) which was theoretically predicted \cite{fCPNernst1}, and experimentally observed in amorphous films of Nb$_x$Si$_{1-x}$ \cite{fCPNernst2} and heavy-fermion superconductor URu$_2$Si$_2$ \cite{fCPNernst3}. Here, we specifically study the temperature-dependence of the thermoelectric coupling strength of two-dimensional (2D) fluctuation Cooper pairs, which we systematically compare to that of charged fermions and bosons in the normal phase. 

\paragraph*{Thermodynamic analysis} Consider a system, like a reservoir, composed of $N$ noninteracting charge carriers with a given statistics at thermal equilibrium at temperature $T$. From the assumption of extensivity of the free energy, one obtains the Gibbs-Duhem relationship: $S{\rm d}T + N{\rm d}\mu = 0$, where $S$ is the system's entropy and $\mu$ the chemical potential. This equality shows that heat and electricity are coupled through the intensive variables $\mu$ and $T$. Now, in analogy with the classical gas, using the correspondence: $V\longrightarrow N$ and $-P\longrightarrow \mu$, we define the following thermoelastic coefficients of the charge carriers as $\beta N = \left(\partial N/\partial T\right)_{\mu}$: analogue to thermal dilatation coefficient; $\chi_T N= \left(\partial N/\partial \mu\right)_{T}$: analogue to isothermal compressibility; $C_{\mu} N= T\left(\partial S/\partial T\right)_{\mu}$: analogue to specific heat at constant pressure; $C_N N= T\left(\partial S/\partial T\right)_{N}$: analogue to specific heat at constant volume. Using extended Maxwell's relations, we find that $\beta/\chi_T = S_N$ with $S_N = \left(\partial S/\partial N\right)_T$, which reflects the notion of entropy per particle introduced by Callen \cite{Callen1} and the ensuing thermodynamic definition of the thermoelectric coupling, $s_{\rm th}=\beta\chi_T^{-1}/q$, which simply shows that the considered particles carry an electric charge $q$ \emph{and} energy.

The relationship between the heat capacities $C_{\mu}$ and $C_N$ is central in the present work:
\begin{equation}\label{CmuCN}
\frac{C_{\mu}}{C_N} = 1 + \frac{\beta^2}{\chi_T C_N}T = 1 + \frac{s_{\rm th}^2}{\ell}= 1 + Z_{\rm th}T
\end{equation}
\noindent where $\ell = C_N/q^2\chi_T T$ is a quantity similar to the Lorenz number \cite{Brantut13}. It is interesting to note the relationship between $\chi_T$ and a capacitance in circuit theory: on the one hand the classical isothermal compressibility is a measure of the change in the system volume as the applied pressure changes; now with the correspondence $V \longrightarrow N$ and $-P \longrightarrow \mu$, we obtain a measure of the ability of a capacitor to store electric charges under an applied voltage, so that $q^2\chi_T$ is an electrical capacitance. Therefore, while the Lorenz number $L$ pertains to coupled transport, the ratio $\ell$ is a quantitative measure of a system's relative ability to store thermal energy with respect to its ability to store electrical charges. The ratio $C_{\mu}/C_N$ is analogous to the classical isentropic expansion factor $\gamma$, and may be used to define the thermodynamic figure of merit, $Z_{\rm th}$, for the charged working fluid. If conditions are found for $C_{\mu}/C_N$ to reach high values or even diverge at \emph{finite} temperature, the electronic working fluid may acquire properties, which could significantly facilitate entropy transport by convection. For a noninteracting many-particle system in the normal phase at equilibrium, the computation of the thermoelastic coefficients $\beta$, $\chi_T$, and $C_{\mu}$ is straightforward (see the Appendix). But, the fluctuation regime necessitates a particular approach~\cite{Larkin,Levin2013}. Calculations detailed in the Appendix yield the chemical potential of the 2D fluctuation Cooper pairs:
\begin{equation}\label{mucp}
\mu_{\rm cp}=\alpha k_{\rm B}T_{\rm c}\epsilon \ln \epsilon
\end{equation}
\noindent where $\epsilon =\ln T/T_{\rm c}\approx (T-T_{\rm c})/T_{\rm c}$ with $T_{\rm c}$ being the critical temperature, $\alpha$ is a dimensionless parameter that enters the definition of the Ginzburg-Landau free energy functional \cite{Larkin}, and $k_{\rm B}$ is the Boltzmann constant. The entropy per particle $s_{\rm th}$ is given by the expression $s_{\rm th} = q^{-1}\partial \mu_{\rm cp}/\partial T$:~\cite{EPLVK}
\begin{equation}\label{sth}
s_{\rm th} = \frac{\alpha k_{\rm B}}{q} \ln\epsilon
\end{equation}
\noindent and the heat capacity ratio reads:
\begin{equation}\label{CmuCNeps}
\frac{C_{\mu}}{C_N} = 1 + \ln\frac{1}{\epsilon}
\end{equation}
\noindent It is positive and exhibits a logarithmically divergent behaviour as $T\rightarrow T_{\rm c}$. It has a universal character for 2D systems that may be described as fluctuation Cooper pairs. The divergent behaviour of $C_{\mu}/C_N$ contrasts with the standard textbook cases of the ratio $C_P/C_V$ for classical and quantum gases in the normal phase for which the isentropic expansion factors remain finite at finite temperatures, in each case \cite{Blundell}. 

\paragraph*{Effective thermoelectric coupling}
As the Seebeck coefficient is $s \overrightarrow{\nabla} T= q^{-1}\overrightarrow{\nabla} \mu$, where $\overrightarrow{\nabla}$ denotes the spatial gradient, it is quite tempting to liken $s$ to $s_{\rm th}$ defined above, using simple dimensional analysis but this would pose some conceptual problems: $s_{\rm th}$, which derives from thermoelastic coefficients, belongs to the field of thermostatics where spatial gradients and the ensuing out-of-equilibrium situation are meaningless, while $s$ describes a process pertaining to transport theory and irreversible processes \cite{deGroot}, where forces and fluxes are meaningful \cite{Onsager1}. In fact, both $s$, which is the degree of mutual interaction at the local level between two irreversible processes \cite{KedemCaplan}, and $s_{\rm th}$, which is the average at the macroscopic level of the system's entropy distributed over its constituents \cite{Callen1}, combine and permit the thermoelectric transport. More precisely, if one considers a simple setup made of two reservoirs and a conducting channel \cite{bookMeso1,bookMeso2}, it is easy to show \cite{Brantut13} that the effective thermoelectric coupling $s_{\rm eff}$, combines the thermodynamic properties of the reservoirs through $s_{\rm th}$ with those of the channel through $s$, as $s_{\rm eff}=s_{\rm th} - s$. As noted in Ref.~\cite{Brantut13}, the effective thermoelectric coupling characterizes two processes: the transport of entropy through the channel as a response to an external constraint (see the Appendix), and the production of entropy as one charge is released from one reservoir and is absorbed by the other. 

\paragraph*{Numerical results and analysis}
To compare \emph{qualitatively} the thermodynamic figure of merit $Z_{\rm th}$ and the thermoelectric heat capacity ratio $C_{\mu}/C_N$ of a system of fluctuation Cooper pairs, to those of the standard 2D and 3D Bose gases, and of the 2D Fermi gas in the normal state, each of these latter being characterized by its equilibrium distribution, chemical potential, and density of states, we compute numerically their thermoelastic coefficients. For simplicity, we adopt the same notations for all the considered systems, and more importantly we base our discussion on the temperature- or equivalently, the $\epsilon$-dependence of $Z_{\rm th}$ and $C_{\mu}/C_N$. The 2D Fermi and Bose systems do not undergo a phase transition, while the 3D Bose gas does at the condensation temperature $T_{\rm cond}$. This permits the definition of a common parameter $\epsilon = \ln (T/T_{\rm cond})$ for the joint analysis of the Bose gases and the Fermi gas; to account for the fluctuation Cooper pairs with a critical temperature $T_{\rm c}$ different from $T_{\rm cond}$, we use the same scale as for the three other systems.

\begin{figure}
	\centering
		\includegraphics[width=0.5\textwidth]{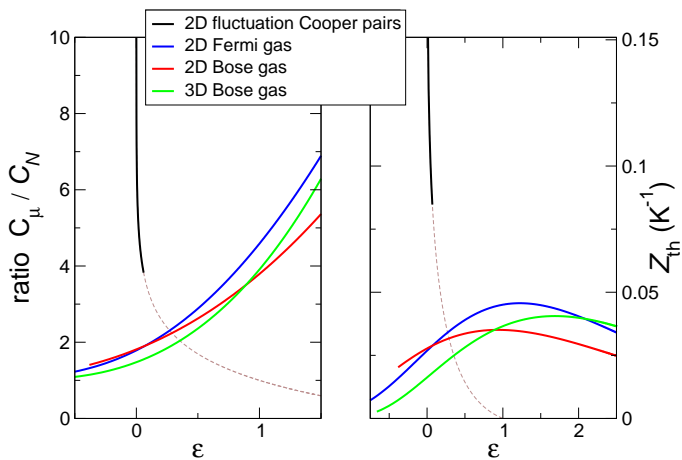}
	\caption{Thermoelectric heat capacity ratio $C_{\mu}/C_N$ and thermodynamic figure of merit $Z_{\rm th}$ as functions of $\epsilon$ for various systems. For the numerical illustration of our analysis, we assume that the Bose and Fermi particles' masses equal that of the free electron, except for the Cooper pairs whose mass is twice as large. The particle concentrations are $n=10^{12}$ cm$^{-2}$ and $n=10^{18}$ cm$^{-3}$ for the 2D and 3D systems respectively. The presence of the dashed lines on both panels simply reflects the logarithmic behaviours of $C_{\mu}/C_N$ and $Z_{\rm th}$, but outside the limit of validity of the model ($\epsilon \ll 1$), so they have no physical meaning; for the fluctuation Cooper pairs, only the black parts of the curves are relevant and meaningful. The $\epsilon$-dependences of $C_{\mu}/C_N$ and $Z_{\rm th}$ are, as expected, essentially the same for the Bose gases and the Fermi gas: they converge towards a finite value as $\epsilon \rightarrow 0$. Unlike the ratio $C_{\mu}/C_N$, $Z_{\rm th}$ has a maximum: as $T$ increases, the entropy per particle $qs_{\rm th}$ can only increase for a fixed number of particles as thermal energy is put into the system; but as $Z_{\rm th}$ also contains a $1/T$ term, it is the rapidly increasing product $\ell T$, which lowers $Z_{\rm th}$ in the high temperature regime. This is in stark contrast with the diverging behaviour near $\epsilon = 0$, i.e. at \emph{finite} temperature for the fluctuation Coopers.}
	\label{fig:figure1}
\end{figure}

Focusing first on the two Bose gases and the Fermi gas, we see on Fig.~\ref{fig:figure1}, that as their temperatures increase, the ratios $C_{\mu}/C_N$ increase monotonically while the figures of merit $Z_{\rm th}$ do not: they reach a maximum value for each particular case and then decrease. This behaviour observed for the ratio $C_{\mu}/C_N$ is akin to that of the ratio $C_P/C_V$ discussed in textbooks, and is of limited interest here since the increase results from the rise of temperature. We also see that as $Z_{\rm th} = s_{\rm th}^2/\ell T$ decreases after reaching its maximum, a temperature increase cannot guarantee optimal conditions for the working fluid. More precisely, for the two Bose gases $\ell$ increases but more slowly than $q^2s_{\rm th}^2$ does, and it saturates, which means that the capacity at which the system may store thermal energy dominates its capacity to store bosonic charges as the temperature increases up to a certain point. For the Fermi system $\ell$ decreases, and in this case, it is the capacity of the system to store fermionic charges that dominates its capacity to store thermal energy. This may appear as a counter-intuitive fact due to the Pauli-blocking mechanism, but it is precisely this latter which permits this: for a given particle number, the Bose distribution falls off more rapidly than the Fermi distribution does and it even gets smaller from a point which depends on the temperature, and as shown by its definition, the electrical capacitance $q^2\chi_T$ follows essentially the behaviour of the statistical distribution to which it is related. Now, unlike the three cases discussed above, the range over which $C_{\mu}/C_N$ and $Z_{\rm th}$ vary for the fluctuation Cooper pairs is restricted to $0<\epsilon \ll 1$. The key point here is that the thermodynamic figure of merit \emph{and} the thermoelectric heat capacity ratio behave in the same fashion: both diverge as $T$ approaches $T_{\rm c}$, which is a finite temperature. This is precisely the desired behaviour, which ensures that the working fluid's thermoelastic properties are optimal, but this is possible only at the cost of very specific conditions on the system's temperature.

\paragraph*{Discussion and concluding remarks}
We showed with the illustrative case of 2D fluctuation Cooper pairs that approaching the phase transition of the electronic working fluid provides the best and ultimate way of significant performance increase, as the thermoelectric coupling as well as the heat capacity ratio and the thermodynamic figure of merit show a divergent behaviour as $T\rightarrow T_{\rm c}$. That the heat capacity ratio may diverge implies that $C_{\mu}$ increases faster than $C_N$ does for the 2D fluctuation Cooper pairs as the system approaches the critical temperature. In classical thermodynamics, the heat capacity at constant pressure $C_P$ may increase while the heat capacity at constant volume $C_V$ may not, the more compressible a fluid is. Therefore, the electronic systems of interest for thermoelectric applications are those whose compressibility is sufficiently high so that the thermoelectric heat capacity ratio $C_{\mu}/C_N$ may significantly increase or even diverge at finite temperature. This is consistent with the conclusions of a study of the superconducting properties of carbon nanotube ropes \cite{Alvarez}: as the compressibility increases, the system becomes inhomogeneous as reflected by the density-density correlation function. It is also of interest to note that Eq.~\eqref{CmuCNeps} assumes a very simple form and that it may be applied to any system which may be described with a fluctuation Cooper pair approach. Possible systems include those that allow excitonic BCS-like pairing of two electrons \cite{Laussy1,Laussy2}, and particularly those which couple to light such as quatron-polaritons \cite{Kavokin}. Indeed, owing to their minute effective mass and high critical temperature \cite{Kavokin} an estimation of the parameter $\alpha$ of Eq.~\eqref{sth} for quatrons (in the Appendix) shows it is seven orders of magnitude greater than that of standard 2D fluctuation Cooper pairs. This indicates that one may expect a \emph{huge} fluctuation bosonic thermoelectric effect above the quatron superconducting transition, and that quatrons could also boast a giant thermomagnetic response.

We specialized our study on the thermodynamics of the electronic working fluid rather than the actual transport problem or, equivalently, on the chemical potential rather than the electrochemical potential. It is the difference of this latter between two reservoirs that generates the electromotive force responsible for the electrical current in thermoelectric devices; so inasmuch an electrical potential does not affect the temperature-dependence of the thermoelastic properties of the working fluid, we could essentially concentrate on their optimization, the necessity of which has been clearly demonstrated with the temperature-dependence of $C_{\mu}/C_N$ and that of $Z_{\rm th}$ for different systems. That $Z_{\rm th}$ may diverge in one case, but not for the standard three others (irrespective of statistics and dimensionality) thus explains why after decades of intense efforts to improve their performance, thermoelectric devices still remain poorly efficient energy-conversion devices. The fluctuation regime studied in this work, where either phonons or excitons are put to work to bind electrons, illustrates the actual possibility to prepare highly compressible electrically charged working fluids; but electronic systems in different configurations to be found, could also present enhanced thermoelectric properties as long as they boast a high compressibility factor. We thus wish to stimulate experimental activities in this genuinely promising new direction. 

\section*{Appendix}

\subsection*{Thermoelastic coefficients in the fluctuation regime}
Since we are principally interested in the ratio $C_{\mu}/C_N$, we may obtain it from the knowledge of the free energy of the 2D fluctuation Cooper pairs, $\mathcal{F}_{\rm cp}$. From Eq.~(2.60) of Ref.~\cite{Larkin}, we get:
\begin{equation}
\mathcal{F}_{\rm cp}=-\frac{{\mathcal A}}{4\pi \xi^{2}}k_{\rm B}T_{\rm c}~\epsilon\ln \epsilon
\end{equation}
\noindent where $\epsilon =\ln T/T_{\rm c}\approx (T-T_{\rm c})/T_{\rm c}$ with $T_{\rm c}$ being the critical temperature, ${\mathcal A}$ is the surface area of the sample containing the Cooper pairs, $k_{\rm B}$ is the Boltzmann constant, and $\xi$ is the coherence length. Then, from Eq. (2.151) of Ref.~\cite{Larkin}, assuming a zero anisotropy parameter and a single superconducting layer we get $N_{\rm cp}=-{\mathcal A}(4\pi \alpha \xi^{2})^{-1}\ln \epsilon$, where $\alpha = \hbar^2/(2m_{\rm cp}k_{\rm B}T_{\rm c}\xi^2)$ is a dimensionless parameter that enters the definition of Ginzburg Landau free energy functional \cite{Larkin}, with $m_{\rm cp}$ being the Cooper pair mass. It thus follows that the chemical potential of the 2D fluctuation Cooper pairs reads:
\begin{equation}
\mu_{\rm cp}=\frac{\partial \mathcal{F}_{\rm cp}}{\partial \epsilon } \times \left(\frac{\partial N_{\rm cp}}{\partial \epsilon }\right)^{\!\!-1}=\alpha k_{\rm B}T_{\rm c}\epsilon \ln \epsilon
\end{equation}
\noindent which constitutes a simple analytic expression with the peculiarity that it does not exhibit an explicit dependence on the number of fluctuation Cooper pairs. The heat capacity at constant particle number is $C_N = (-T/N_{\rm cp}) \partial^2\mathcal{F}_{\rm cp}/\partial T^2$:
\begin{equation}
C_N N_{\rm cp}= \frac{{\mathcal A}}{4\pi\xi^2}k_{\rm B}\left(\ln\epsilon + \frac{1}{\epsilon} +1\right) \approx \frac{{\mathcal A}}{4\pi\xi^2\epsilon}k_{\rm B}
\end{equation}
\noindent since $1/\epsilon$ is the dominant term as $T$ approaches $T_{\rm c}$. The analogue to the thermal compressibility $\chi_T$ is: 
\begin{equation}
\chi_T N_{\rm cp} \!\!=\!\! \left(\frac{\partial N_{\rm cp}}{\partial \mu_{\rm cp}}\right)_{\!\!T}\!\! =\!\! \frac{\partial N_{\rm cp}}{\partial \epsilon}\left(\frac{\partial  \mu_{\rm cp}}{\partial \epsilon}\right)^{\!\!-1}\!\!=\!\!\frac{-{\mathcal A}}{4\pi\alpha^2\xi^2 k_{\rm B}T_{\rm c}} \frac{1}{\epsilon \ln\epsilon}\\
\end{equation}
\noindent So, the ratio $\ell = C_N/q^2\chi_T T$ explicitly reads: 
\begin{equation}
\ell = - \frac{\alpha^2}{q^2}k_{\rm B}^2\ln\epsilon
\end{equation}

\subsection*{Thermoelastic coefficients in the normal phase} 
The thermoelastic coefficients $\beta$, $\chi_T$, and $C_{\mu}$ were given in Ref.~\cite{Brantut13}. For clarity we give the derivation steps for $\chi_T$; the other two follow similar steps. For a many-particle system with energy distribution function $f$ and density of state $g$, the number of particles $N$ is given by:

\begin{equation}
N = \int_0^{\infty} g(E)f(E){\rm d}E
\end{equation}

\noindent By definition, $\chi_T N= \left(\partial N/\partial \mu\right)_{T}$, and the first partial derivative of $N$ with respect to $\mu$ at constant temperature takes the form:

\begin{equation}
\frac{\partial N}{\partial \mu} = \int_0^{\infty} g(E)\frac{\partial f}{\partial \mu}{\rm d}E = \int_0^{\infty} g(E)\left(-\frac{\partial f}{\partial E}\right){\rm d}E
\end{equation} 

\noindent since the density of state does not depend on $\mu$ and $\partial f/\partial \mu = -\partial f/\partial E$. We thus obtain $\chi_T$ as well as the other two coefficients in the same fashion:

\begin{eqnarray}
\nonumber
\chi_T N&=& \int_0^{\infty} g(E)\left(-\frac{\partial f}{\partial E}\right) {\rm d}E\\
\nonumber
\beta N&=& \frac{1}{T_{\rm a}}\int_0^{\infty} g(E)\left(E-\mu_{\rm a}\right)\left(-\frac{\partial f}{\partial E}\right) {\rm d}E\\
\nonumber
C_{\mu} N&=& \frac{1}{T_{\rm a}}\int_0^{\infty} g(E)\left(E-\mu_{\rm a}\right)^2\left(-\frac{\partial f}{\partial E}\right) {\rm d}E
\end{eqnarray}

\subsection*{Seebeck coefficient and entropy transport}
For bosonic and fermionic systems in the normal phase, the Seebeck coefficient for the channel reads \cite{Mahan}: $s = L_{12}/L_{11} = k_{\rm B}{\mathcal K}_1/{\mathcal K}_0$, with the $L_{ij}$'s being the kinetic coefficients computed from the kinetic integrals ${\mathcal K}_n = \int \left(\frac{E-\mu}{k_{\rm B}T}\right)^n\Sigma(E)\left(-\frac{\partial f}{\partial E}\right) {\rm d}E$ where $\Sigma(E)$ is the transport distribution function of the channel, which for simplicity assumes the standard Lorentzian shape. The Seebeck coefficient may thus be seen as the average value of the entropy involved in the thermoelectric transport, $(E-\mu)/T$, over a probability density function given by the product of the transmission function and the energy derivative of $f$. Note that in metals and degenerate semiconductors, as the electrons above the Fermi level carry a heat current that is practically the opposite of that carried by the electrons below the Fermi level, the Seebeck coefficient is typically small; in other words, from the mathematical viewpoint, since $E-\mu$ changes sign as $E$ varies, it is essential that $\Sigma$ presents an asymmetric profile~\cite{Adel}, to avoid cancellation of ${\mathcal K}_1$. It is easy to check numerically for bosonic and fermionic systems in their normal phase, that $s$ decreases and tends to 0 as the system's temperature keeps decreasing. Turning to the fluctuation regime, the general derivation of the Seebeck coefficient for the two-dimensional fluctuation Cooper pairs is much more involved~\cite{Larkin}, but it yields $s \sim \ln(1/\epsilon)$ as $T \rightarrow T_{\rm c}$.

\subsection*{Quatron-polaritons in the fluctuation regime}
Quatrons are composite quasi-particles, formed of two electrons located in separated quantum wells but bound by an attractive effective Coulomb interaction mediated by an exciton located in a third quantum well sandwiched between the other two. Within an optical microcavity, quatron form quatron-polaritons (QP) thanks to the coupling of the electronic excitations to the photonic modes of the cavity; their effective mass is typically $10^4$ times smaller than that of free electrons. Under some experimental conditions, in a suitably designed structure, a system of QP may undergo a superfluid phase transition near or at room temperature owing to the extremely light QP effective mass. If the quatrons form a working fluid, which is prepared near its phase transition, its thermodynamic figure of merit $Z_{\rm th}$ and its thermoelectric heat capacity ratio $C_{\mu}/C_N$ are given by Eq.~\eqref{CmuCNeps} with the same $\epsilon$-dependence as for the standard Cooper pairs. It is of interest to note that due to the QP's minute effective mass, their critical temperature $T_{\rm c}$ is much higher than that of standard superconducting transition. A crucial difference will also be in the thermoelectric coupling $s_{\rm th}$, which explicitly depends on the parameter $\alpha$ as shown in Eq.~\eqref{sth}. Now, let us compare the values taken by the parameter $\alpha$ for quatrons, $\alpha_{\rm q}$, with that for the standard 2D Cooper pairs, with mass $m_{\rm cp}$, which in the phenomenological theory of Ginzburg and Landau, explicitly reads \cite{Larkin}:

\begin{equation}
\alpha_{\rm GL}^{\rm 2D}=\frac{4\pi^{2}k_{\rm B}T_{\rm c}}{7\zeta(3)E_{\rm F}}
\end{equation}

\noindent where $E_{\rm F}$ is the Fermi energy. The parameter $\alpha_{\rm GL}^{\rm 2D}$ is very small. For quatrons the mass $m_{\rm q}\sim 10^{-4} m_{\rm e},$ while $\xi\ll \xi_{GL}^{\rm 2D}$, i.e., typically two orders of magnitude smaller than the superconducting coherence length. Then,

\begin{eqnarray}
\nonumber
\alpha_{\rm q} &=& \frac{1}{2m_{\rm q}T_{\rm c,q}\xi^2}=\frac{m_{\rm cp}}{m_{\rm q}}\times\frac{T_{\rm c,GL}}{T_{\rm c,q}}\left(\frac{\xi_{GL}^{\rm 2D}}{\xi}\right)^2 \alpha_{\rm GL}^{\rm 2D} \\
{}&~&\sim 10^{8}\alpha_{\rm GL}^{\rm 2D}\left(\frac{T_{\rm c,GL}}{T_{\rm c,q}}\right)
\end{eqnarray}

\noindent Under the reasonable assumption that $T_{\rm c,q}$ would be of the order of room temperature, $T_{\rm c,GL}/T_{\rm c,q}\sim 10^{-1}$ and 

\begin{equation}
\alpha_{\rm q} \sim 10^7 \alpha_{\rm GL}^{\rm 2D}
\end{equation}

\end{document}